\documentclass[letterpaper,twocolumn,10pt]{article}
\usepackage{usenix,graphicx,endnotes}
\usepackage{enumitem}
\usepackage{balance}
\usepackage[available]{usenixbadges}
\begin{document}

\date{}

\title{\Large \bf The Impact of Security and Privacy Controls on Users' Emotional Engagement with Generative AI Chatbots}

\pagestyle{empty}

\author{
{\rm Jabari Kwesi}\\
Duke University
\and
{\rm Jiaxun Cao}\\
Duke University
\and
{\rm Hailee Cunningham}\\
Duke Kunshan University
\and
{\rm Pardis Emami-Naeini}\\
Duke University
}

\maketitle

\subsection*{Abstract}

Chatbots powered by generative AI (e.g., OpenAI's ChatGPT and Google's Gemini) are increasingly being appropriated for emotional support and companionship. These tools offer a suite of security and privacy (S\&P) controls, including model training opt-outs and memory toggles, yet how the presence of these controls influences users' attitudes toward emotionally sensitive disclosure remains understudied. We conducted a mixed-methods vignette study with 354 U.S. participants to examine how S\&P controls influence users' willingness to engage with generative AI chatbots for emotional support, their perceptions of how protected they are when using these systems, and their perceptions of how effective the chatbots are for providing support. Controls enabling deletion of disclosures had the largest positive impact: these offerings outperformed technically sophisticated controls such as local-only processing and model training opt-outs, where participants expressed difficulty understanding the underlying mechanisms. Yet trust remains fragile, and participants often doubted S\&P controls would function as promised. We conclude with actionable recommendations informed by our results to bridge users' comprehension gaps, build credible assurances, and properly calibrate barriers for users in distress.

\section{Introduction}
\label{sec:introduction}
Digital interventions for emotional support have historically been dominated by rule-based systems designed to guide users through structured therapeutic exercises~\cite{abd2019overview}. Though effective for symptom management, these systems' rigid architecture limits personalization and fails to handle open-ended emotional input~\cite{hua2025charting, yeh2025does}. Users are instead migrating towards platforms enabled by Large Language Models (LLM) for emotional support~\cite{kwesi2025exploring}, such as OpenAI's ChatGPT, Google's Gemini, and Anthropic's Claude, as well as purpose-built AI companions like Luka's Replika and Character.AI; these systems support context-aware dialogue increasingly capable of fostering emotional intimacy~\cite{de2023benefits, ma2023understanding, song2024typing}. Users are rapidly appropriating generative AI (GenAI) systems for health and well-being for which they were not designed, at a scale that dwarfs that of traditional crisis infrastructure; ChatGPT alone fields over one million suicide-related conversations weekly, several times the volume of the U.S. 988 Suicide and Crisis Lifeline~\cite{techcrunch2025suicide, kff2025988}. OpenAI’s January 2026 launch of ChatGPT Health reflects a broader move to formalize these appropriations~\cite{openai2026health}, yet the privacy and safety implications of this scale remain an open question.

The risks of this shift align with well-documented concerns in user disclosures of personal informatics and the quantified self, where intimate data is routinely commodified~\cite{lupton2016diverse}. In this specific instance, users are submitting detailed depictions of emotional turmoil to commercial systems, often without calculating the consequences of sharing sensitive information with platforms that may retain or re-purpose their disclosures~\cite{king2025user}. This phenomenon has been termed \textit{intangible vulnerability}: a risk perception gap where users underestimate the long-term harms of commodifying their emotional inner lives compared to tangible risks like financial fraud~\cite{kwesi2025exploring}. As most of these platforms currently operate outside traditional healthcare regulation~\cite{kwesi2025exploring, rezaeikhonakdar2023ai}, user-facing security and privacy (S\&P) controls are often the only mechanisms through which users can exercise agency over their data. Prior work has documented the technical risks of these systems as used for mental health and emotional support, as well as the cognitive limitations users bring to privacy decisions, yet we know little about how users perceive these controls or whether they influence willingness to engage.

To address these gaps, our study explores three primary research questions:

\begin{itemize}
    \item \textbf{RQ1:} How do the S\&P controls available in GenAI chatbots influence users' \textbf{willingness to use} the chatbot for emotional disclosures?
    \item \textbf{RQ2:} How do the S\&P controls available in GenAI chatbots influence users' \textbf{perceptions of protection} provided by the chatbot when sharing emotional disclosures?
    \item \textbf{RQ3:} How do the S\&P controls available in GenAI chatbots influence users' \textbf{perceived efficacy} of the chatbot for emotional support?
\end{itemize}

To answer these questions, we conducted a systematic audit of 87 chatbots enabled by generative AI to derive a taxonomy of nine representative S\&P controls. We deployed a mixed-methods vignette study ($N=354$) in which U.S.-based participants rated the influence of these S\&P controls on their \emph{willingness to engage}, \emph{perceived protection}, and \emph{perceived efficacy} across three contexts: (1) Anxiety, (2) Depression, and (3) Interpersonal Tension. To complement these quantitative insights, we collected and analyzed written explanations of participants' reasoning.

Our statistical findings showed that disclosure context had no significant effect on user perceptions: patterns held across anxiety, depression, and interpersonal tension. Deletion controls dominated user preferences, though participants expressed skepticism about whether such controls could fully remove their information or the inferences derived from their conversations. Most S\&P controls showed similar effects across willingness, perceived protection, and perceived efficacy, with notable exceptions: \emph{Multifactor Authentication (MFA)}, for instance, was recognized as protective but reduced willingness to engage due to friction in moments of vulnerability. We propose actionable design recommendations to address the three axes on which our findings revealed participants struggled the most: comprehension, assurance, and \textit{affective urgency}. The first captures struggles to understand technically sophisticated controls; the second reflects skepticism that controls will function as promised; the third describes tension between protective friction and accessibility during emotional distress. We then situate these recommendations as assessment criteria, amidst a discussion of federal and state policymakers' recent divergence on the regulation of AI in sensitive and emotional contexts.

\section{Background \& Related Work}

\subsection{Conversational AI in Emotional Support}

A growing body of research, including over 35 studies synthesized in Li et al.'s recent meta-analysis, establishes the therapeutic potential of conversational agents for mental health. Early randomized controlled trials have demonstrated that rule-based chatbots can reduce symptoms of depression and anxiety~\cite{fitzpatrick2017delivering, inkster2018empathy}; the aforementioned meta-analysis found moderate-to-large effects on psychological distress and depression, with notable efficacy being attributed to generative agents~\cite{li2023systematic}.

These findings have catalyzed a shift from purpose-built agents towards general-purpose Large Language Models (LLMs). Where rule-based systems have been found by users to be rigid and impersonal~\cite{burton2016pilot, fitzpatrick2017delivering, huang2020challenges}, LLMs enable open-ended dialogue that adapts to nuanced emotional input~\cite{kirk2024benefits, ma2023understanding}. Users report finding these systems more context-aware and less ``on-rails'' than their rule-based predecessors~\cite{song2024typing, sharma2023cognitive}. 

As a result, general-purpose AI systems are increasingly being appropriated for emotional labor at a scale that dwarfs traditional crisis infrastructure. ChatGPT receives over 800 million weekly visitors, hundreds of millions of whom ask questions related to health and wellness~\cite{openai2026health, fortune2026health}. OpenAI estimates that over one million of these users discuss suicide each week, in conversations exhibiting explicit indicators of suicidal planning or intent~\cite{techcrunch2025suicide}. This appropriation occurs alongside rapid proliferation of purpose-built AI companions: these are platforms like Replika and Character.AI, explicitly designed to foster emotional intimacy and ``unidirectional bonds''~\cite{skjuve2021my, pentina2023exploring}. These bonds can become pathological; Laestadius et al.'s grounded theory analysis found emotional dependence resembling unhealthy human relationships~\cite{laestadius2024too}.

These systems require continuous and high-fidelity self-disclosure to function effectively as an emotional supplement. Lee et al.'s experimental work confirms this structure is reciprocal: chatbot self-disclosure both promotes deeper user disclosure on sensitive topics and enhances perceived conversational intimacy~\cite{lee2020selfdisclosure}. This creates a structural tension in which users must perceive the agent as a confidant deserving of intimate disclosure, while backend architectures process these disclosures as training data unless otherwise specified by the user~\cite{king2025user}. Emotional disclosures are particularly worrisome, as they can reveal mental state, behavioral patterns, location-based info, and psychological vulnerabilities through subtle linguistic markers~\cite{ngong2025protecting, zhang2024sa}, yet users undervalue these risks relative to financial or biometric data in what prior work terms ``intangible vulnerability''~\cite{kwesi2025exploring}. Gaps between user mental models and \textit{de facto} system operation deepen this undervaluation: many users believe their chatbot interactions are protected by extant regulatory frameworks (\textit{e.g.,} HIPAA), despite most platforms operating outside such protections~\cite{kwesi2025exploring, wang2025mental}. In practice, emotional disclosures are liable to be incorporated into training data, where language models can memorize them at elevated rates and render them vulnerable to extraction~\cite{carlini2021extracting, du2025beyond}.

\subsection{Risks and Challenges}

The rapid transition to generative AI for emotional support introduces distinct risks.  At a technical level, LLMs suffer from well-documented reliability issues: these include the production of inaccurate or iatrogenic information (hallucinations), biased content generation, limited interpretability, and general inconsistency~\cite{weidinger2021ethical, guo2024large, ji2023rethinking}. These systems are also capable of inferring sensitive personal attributes from seemingly innocuous text with up to 85\% accuracy---at \textit{100x} lower cost and \textit{240x} greater speed than human profilers, and even when users attempt to anonymize their language~\cite{staab2024beyond, shanmugarasa2025sok}. Users routinely disclose the kind of sensitive information, namely spatiotemporal information and PII, that enables such inference~\cite{mireshghallah2023can}; even privacy-conscious users inadvertently leak sensitive context during high-affect interactions~\cite{ngong2025protecting}. Recent work suggests that the personalization making AI increasingly effective in emotional use cases also amplifies these risks, as adapting to individual users requires building profiles that create vectors for privacy infringement and bias reinforcement~\cite{kirk2024benefits}.

These technical risks are compounded by documented limitations in privacy decision-making, especially in sensitive contexts as emotional support. Decades of privacy research establish that users systematically underestimate privacy risks when disclosing sensitive information~\cite{acquisti2015privacy, norberg2007privacy, acquisti2005privacy}, in a pattern Acquisti's foundational work attributes to \textit{bounded rationality} and \textit{hyperbolic discounting} of future harms~\cite{acquisti2004privacy}. Privacy expectations are almost entirely contextual per Nissenbaum's seminal \textit{contextual integrity}, and information flows acceptable in one context become violations in another~\cite{nissenbaum2004contextual}. Counterintuitively, providing users with more control over their data can \textit{increase} disclosure risk rather than protect them, in what has been identified as the ``control paradox''~\cite{brandimarte2013misplaced}.

Both the technical vulnerabilities and the behavioral limitations they exploit exist within what is currently a regulatory vacuum. The FDA has authorized over 1,200 AI-enabled devices, but none for mental health applications; purpose-built chatbots like Woebot and Wysa have received breakthrough device designations but not approval~\cite{fdabreakthrough2022}. General-purpose LLMs face no FDA oversight at all because they make no health claims, even as millions appropriate them for emotional support. State responses are emerging but fragmented: California now mandates suicide prevention protocols and disclosure for AI chatbots~\cite{casb243}, Illinois and Nevada have banned AI from providing mental health treatment~\cite{illinoiswopr2025, nevadaab406}, and New York requires notifications for emotionally-responsive AI companions. Yet these laws primarily target platforms that \textit{claim} to provide mental health services or licensed professionals who deploy them, while leaving the appropriation of general-purpose tools largely unaddressed. The consequences of this vacuum are already arriving: Character.AI settled the first wave of lawsuits alleging that AI companion chatbots contributed to teen suicides~\cite{characterai2026settlement} in January 2026, and peer support communities have already emerged to assist users affected by what members describe as ``AI delusions or spirals''~\cite{npr2026humanline}. Yet legal frameworks remain unsettled, and in this vacuum, user-facing UI controls may represent the primary mechanism through which users can exercise agency over their safety---though we know little about how users perceive these controls. Of note is Das et al.'s systematic review of multi-factor authentication research, wherein the authors found that fewer than 10\% of recent studies included any user evaluation~\cite{das2019evaluating}---those studies that did consult users documented persistent adoption challenges, even when users recognized protective value. Whether similar dynamics apply to perception of S\&P controls in emotional contexts remains unexplored. This study provides what is to our knowledge the first systematic evaluation of how user-facing S\&P controls influence perceptions of protection, efficacy, and willingness to engage among users who appropriate GenAI chatbots for emotional support.

\section{Methodology} \label{methods}
To identify the currently available S\&P controls in GenAI chatbots, we conducted a systematic audit of 87 consumer apps. We then launched a mixed-methods survey ($N=354$) on the Prolific crowdsourcing platform~\cite{palan2018prolific} to capture participants' attitudes and behaviors toward the surfaced controls. 

\subsection{S\&P Control Audit Procedure}
We conducted a systematic audit of consumer-facing GenAI chatbots. Following established precedents in published human-computer interaction (HCI) work~\cite{lyngs2019self, stawarz2015beyond}, we queried Apple's App Store between August--September 2025 using a set of iteratively refined keywords informed by prior work on conversational agent nomenclature~\cite{zheng2022ux, wang2023investigating}: \emph{Conversational Agent}, \emph{Conversational Interface}, \emph{Chatbot}, \emph{Virtual Agent}, \emph{Virtual Human}, \emph{AI Assistant}, \emph{AI Bot}, \emph{Social Agent}, \emph{Dialogue System}, and \emph{Dialog System}. Initial screening produced 568 listings. For apps with web/desktop clients, we audited all platforms to capture cross-platform differences. In a similar fashion to prior work~\cite{shen2015finding, lyngs2019self}, we manually screened the metadata for each app. We included apps featuring bidirectional AI-driven dialogue (powered by NLP or generative AI) with text or voice interaction, excluding static or purely rule-based bots without dynamic dialogue. We further limited our corpus to English-language apps updated since January 2022 with at least 50 reviews and free-tier availability, and removed duplicates and platform variants. After applying inclusion/exclusion criteria, our working corpus contained 344 distinct apps.

\boldpartitle{S\&P control extraction} In place of auditing all 344 apps, we employed a saturation-based sampling strategy. Two researchers first audited the top 10\% of apps by popularity ($n=34$), reasoning that market leaders would exhibit the most comprehensive S\&P feature sets. We excluded the premium-only controls to focus on users' baseline privacy and security experiences. We then audited three consecutive random 5\% batches from the remaining corpus: the first batch yielded one additional feature. The second and third batches yielded no novel controls, thus confirming taxonomic saturation~\cite{guest2020simple}. In total, we audited 87 apps; the complete audit is available from the authors upon request. The audit produced a nine-feature taxonomy of user-facing S\&P controls, summarized in Table~\ref{tab:taxonomy}. For controls that appeared sparsely across platforms, we additionally aligned their definitions with modern terminology defined by technical literature: MFA follows guidelines established by the National Institute of Standards and Technology~\cite{nist80063b4}, and local-only processing reflects standard academic definitions~\cite{zhou2019edge, xu2024device}. These features served as the vignette stimuli in our survey.

\begin{table*}[t]
\centering

\small
\begin{tabular}{@{}ll@{}}
\toprule
\textbf{S\&P Controls} & \textbf{Operational Definition} \\
\midrule
Delete Conversation & User can permanently delete selected past conversations from account history \\
Delete Account \& Data & User can permanently delete account and all stored data \\
Anonymous Chat & User can control whether a conversation is saved to account history \\
Non-mandatory Login & User can use the chatbot without creating or signing into an account \\
Memory Toggle & User can turn cross-session memory on/off so the chatbot can/cannot recall past conversations \\
Model Training Opt-out & User can control whether conversations are used to train the AI model \\
Access/Sharing Controls & User can control whether conversations are private or shareable publicly \\
Multifactor Authentication (MFA) & User can require additional identification (e.g., SMS, biometric) beyond password \\
Local-only Processing & User can control whether AI computations run locally on-device or via cloud \\
\bottomrule
\end{tabular}
\caption{Nine user-facing S\&P controls derived from a systematic audit of 87 generative AI chatbot applications. Operational definitions reflect common user-facing language observed in free-tier app interfaces.}
\label{tab:taxonomy}
\end{table*}

\boldpartitle{Coder agreement and trustworthiness} In line with HCI guidance that IRR statistics are not always appropriate or necessary for inductive qualitative analyses~\cite{mcdonald2019reliability}, we used a team-based consensus (negotiated coding) approach in place of computing IRR coefficients~\cite{ortloff2023different}. Two researchers collaboratively coded the dataset and met regularly to discuss and resolve discrepancies. 

\subsection{Survey Design \& Procedure}
\label{sec:survey}

\boldpartitle{Participant screening \& eligibility} We used Prolific~\cite{palan2018prolific} to recruit English-speaking U.S. adults (18+). To be eligible, participants were required to use a GenAI chatbot for emotional and/or social support at least monthly. All participants who completed the one-minute screening survey received \$0.20 compensation via Prolific. We provide the study materials, including the recruitment text, screening survey, main survey, and consent form in our supplementary repository (Appendix ~\ref{app:supplementary}).

Prior to the main study, we conducted a pilot survey ($n=15$) with participants recruited via Prolific to both calibrate survey timing and inform power analysis. We revised the wording of several questions for clarity based on pilot feedback. Median completion time was approximately 17 minutes, and a simulation-based power analysis~\cite{green2016simr} indicated adequate sample size for detecting small-to-medium effects in our repeated-measures design at 80\% power. Based on the power analysis, we recruited \textbf{N = 354} participants via Prolific. We provided all participants with informed consent, and our study protocol was approved by our institution's review board (IRB). Full demographic breakdowns are provided in our supplementary materials (Table ~\ref{sec:demographics-full}).

\boldpartitle{Experimental design} We employed a mixed between-subjects and within-subjects design with \textit{S\&P Control} as the within-subjects factor (e.g., Delete Conversation) and the \textit{Context of Disclosure} (e.g., \emph{Anxiety \& Stress}) as the between-subjects factor. To mitigate survey fatigue~\cite{backor2007estimating} and based on our pilot data, each participant evaluated 4 of 9 S\&P controls, while \textit{Context of Disclosure} was assigned from the participant's highest-rated GenAI use case in a pre-task (see Appendix~\ref{app:supplementary}); ties were randomized and quotas maintained balance across contexts. Vignette order was randomized for every participant.

\boldpartitle{Vignette structure} Each vignette instantiated exactly one S\&P control from our taxonomy (Table~\ref{tab:taxonomy}), with a definition directly informed by the language used by the reviewed chatbots. Below is an example vignette that we presented to participants. The italicized words are the factors we varied across scenarios. 

\begin{quote}
    Imagine you are considering using a new AI-enabled chatbot for a conversation to support you through feelings of \emph{Anxiety \& Stress}. Assume the following control is available to you exactly as described. 
    
    \emph{Model Training Opt-out: You can control whether or not your conversations are used to train the chatbot's AI model.}
\end{quote}

\boldpartitle{Survey questions} After each vignette, participants answered three 7-point rating questions to capture the impact of the presented S\&P control on respondents' willingness to use the chatbot, perception of protection offered by the chatbot, and the perceived efficacy of the chatbot for emotional support. These questions were developed specifically for this study to capture directional impact. For example, we asked: \emph{How, if at all, would your willingness to use this chatbot for your conversations about \emph{Anxiety \& Stress} change compared to not having this control?} Options ranged from \emph{Strongly decrease} to \emph{Strongly increase}, centered at \emph{Neither decrease nor increase}. An open-ended prompt followed each rating question to capture participants' rationales.

\boldpartitle{Survey procedure} Participants first provided informed consent and rated their likelihood of using a GenAI chatbot for the three tested contexts (\emph{Anxiety \& Stress}, \emph{Depression \& Low Mood}, \emph{Interpersonal \& Relationship Tension}). These contexts were selected from prior work identifying them to be the most commonly reported presenting concerns in outpatient psychotherapy~\cite{cook2010psychotherapists, perez2017presenting}, and are further supported by recent survey data identifying anxiety, depression, and relationship challenges as three of the four most common motivators for U.S. adults to use LLMs for mental health support~\cite{rousmaniere2025large}. Based on their responses, we determined the \emph{Context of Disclosure} for all presented vignettes by selecting the participant's highest-rated context. Participants then completed four randomized scenarios. After the vignettes and their follow-up questions, we asked participants questions to capture their experience with the S\&P controls they have used in their interactions with GenAI chatbots for emotional support and their general comfort in disclosing emotional information to GenAI compared to a human. We measured participants' baseline comfort disclosing emotional content to GenAI chatbots using items adapted from the Self-Disclosure dimension of Ledbetter's Online Communication Attitude instrument~\cite{ledbetter2009measuring}. We ended the survey by collecting demographic information (e.g., age, gender, race). 

\boldpartitle{Comprehension \& quality checks} Each vignette concluded with a single-best-answer attention check multiple choice question about the described control, in which participants were asked to select the definition of the control they just answered questions about. Options were randomized; only one was correct. ``I'm not sure'' and incorrect responses were scored as fails. Participants failing $\geq$ 2 checks were excluded.

\subsection{Quantitative Analysis}
Because the vignette outcomes are ordered categorical and each participant provides multiple ratings, we fit \emph{cumulative link mixed models} (CLMMs) with a cumulative logit link and a participant-level random intercept~\cite{christensen2025ordinal}. CLMMs allow modeling fixed effects of interest while accounting for dependence between repeated observations~\cite{taylor2023rating}. We utilized a backward elimination~\cite{harrell2015rms} for model selection. Initial models included all experimental factors and demographic information. We used likelihood ratio tests to iteratively prune non-significant predictors that did not improve model fit (based on AIC/LRT~\cite{burnham2002model}). Interaction terms (e.g., \textit{S\&P Control} $\times$ \textit{Context of Disclosure}) were similarly tested and dropped if non-significant. For each dependent variable (DV), below is an example model that we constructed. The formal CLMM specification is provided in Appendix~\ref{sec:clmm-appendix}.

\begin{quote}
\textit{DV} $\sim$ \textit{S\&P Control} + \textit{Context of Disclosure} + \textit{Demographic Factors} + \textit{Disclosure Comfort} + (1 | \textit{Participant})
\end{quote}

Below are the independent variables we included as fixed-effect factors in the model:

\begin{itemize}
\item \textit{S\&P Control} (9 levels): the S\&P control described in the vignette.
\item \textit{Context of Disclosure} (3 levels): \emph{Anxiety \& Stress}; \emph{Depression \& Low Mood}; \emph{Interpersonal \& Relationship Tension}. 
\item \textit{Demographic Factors}: Participants' self-reported race (see Table~\ref{tab:full-clmm} note for covariate selection rationale). 
\item \textit{Disclosure Comfort} (continuous): A standardized 6-item composite ($\alpha = .89$) measuring baseline comfort disclosing emotional content to GenAI chatbots.
\end{itemize}

We report uncorrected $p-values$ for the independent variables across three dependent variables. Benjamini-Hochberg FDR correction confirms that all effects significant at $p < .05$ remain significant after adjustment; Bonferroni correction ($\alpha = .002$) preserves all effects originally reported at $p < .001$.

\boldpartitle{Model baseline selection} For \textit{Context of Disclosure}, \emph{Anxiety \& Stress} was chosen as the baseline category based on alphabetical ordering. For \textit{Control}, we selected \emph{Delete Conversation} as the reference category, as it was the most prevalent feature in our 344-app corpus and represents an industry status quo. Although the choice of baseline does not affect overall model fit or the statistical significance of the predictors, estimated effects should be interpreted relative to their respective baseline categories.

\subsection{Qualitative Analysis}

We analyzed open-ended responses using a codebook-driven thematic analysis~\cite{saldana}. The initial codebook was expanded inductively where new ideas appeared~\cite{saldana}. For each open-ended question, using an iterative open-coding approach~\cite{strauss1990basics}, two researchers independently coded a subset of 10\% of responses and developed the initial codebook inductively. The codebook was refined through team discussion until consensus was reached; disagreements on code distribution were minimal and resolved through verbal deliberation~\cite{ortloff2023different}. The primary coder then analyzed the remaining data. The final codebook comprised 21 primary codes organized around participant mental models, threat perceptions, feature evaluations, and exogenous observations. The full codebook is available in our supplementary repository (Appendix~\ref{sec:codebook}). We performed interpretive synthesis~\cite{barnett2009methods} to map themes to the quantitative findings.

\subsection{Limitations}

Participants evaluating hypothetical S\&P controls may respond differently than users confronting actual S\&P decisions under conditions of emotional distress. Nevertheless, vignette-based designs are an established methodology for enabling systematic comparison across a wide range of conditions (e.g., nine S\&P controls) under standardized circumstances, including controls that do not yet exist on participants' primary platforms. We acknowledge that the S\&P control landscape in generative AI evolves rapidly; our taxonomy reflects the state of consumer apps from August to September 2025. Solove's~\cite{solove2021myth} critique of the privacy paradox and Baruh et al.'s meta-analysis, which found no significant differences between behavioral intentions and actual behavior for privacy-related outcomes across online services and social network use~\cite{baruh2017privacy}, together suggest that stated preferences may correlate 
more reliably with behavior when measured with contextual specificity. Our Prolific sample is limited to U.S.-based participants and skews younger and more educated than the general U.S. population---it also skews towards participants self-identifying as White (65.5\%). Findings may not generalize to other cultural or regulatory contexts. Our screening criterion of monthly GenAI chatbot use for emotional and relationship support necessarily excludes non-users and active avoiders whose privacy concerns may differ systematically. Finally, our control definitions were derived from actual platform language to maximize ecological validity; testing alternative framings may yield different user responses and remains a valuable direction for future work.

\section{Results} \label{sec:results}
\begin{table}[htbp]
\centering
\small
\begin{tabular}{llr}
\toprule[1pt]
\textbf{Category} & \textbf{Value} & \textbf{n (\%)} \\
\midrule
\multirow{2}{*}{Age} & Range & 18--65+ \\
 & Mean (SD) & 43.2 (12.7) \\
\hline
\multirow{3}{*}{Gender} & Female & 211 (59.6\%) \\
 & Male & 132 (37.3\%) \\
 & Non-binary / Other & 11 (3.1\%) \\
\hline
\multirow{5}{*}{Race/Ethnicity} & White & 232 (65.5\%) \\
 & Black or African American & 48 (13.6\%) \\
 & Multiracial / Mixed & 30 (8.5\%) \\
 & Asian & 19 (5.4\%) \\
 & Hispanic or Latino & 15 (4.2\%) \\
\hline
\multirow{4}{*}{Education} & Graduate degree & 77 (21.8\%) \\
 & Bachelor's degree & 114 (32.2\%) \\
 & Some college / Associate & 112 (31.6\%) \\
 & High school or less & 48 (13.6\%) \\
\hline
\multirow{2}{*}{MH Care History} & Ever received care & 216 (61.0\%) \\
 & Currently receiving & 66 (18.6\%) \\
\hline
Tech Background & Yes & 52 (14.7\%) \\
\bottomrule[1pt]
\end{tabular}
\caption{Participant Demographics \textit{(N = 354)}} \label{tab:demographics}
\end{table} 

\begin{figure}[t]
\centering
\includegraphics[trim=.05in .05in .05in .2in, clip, width=0.99\columnwidth]{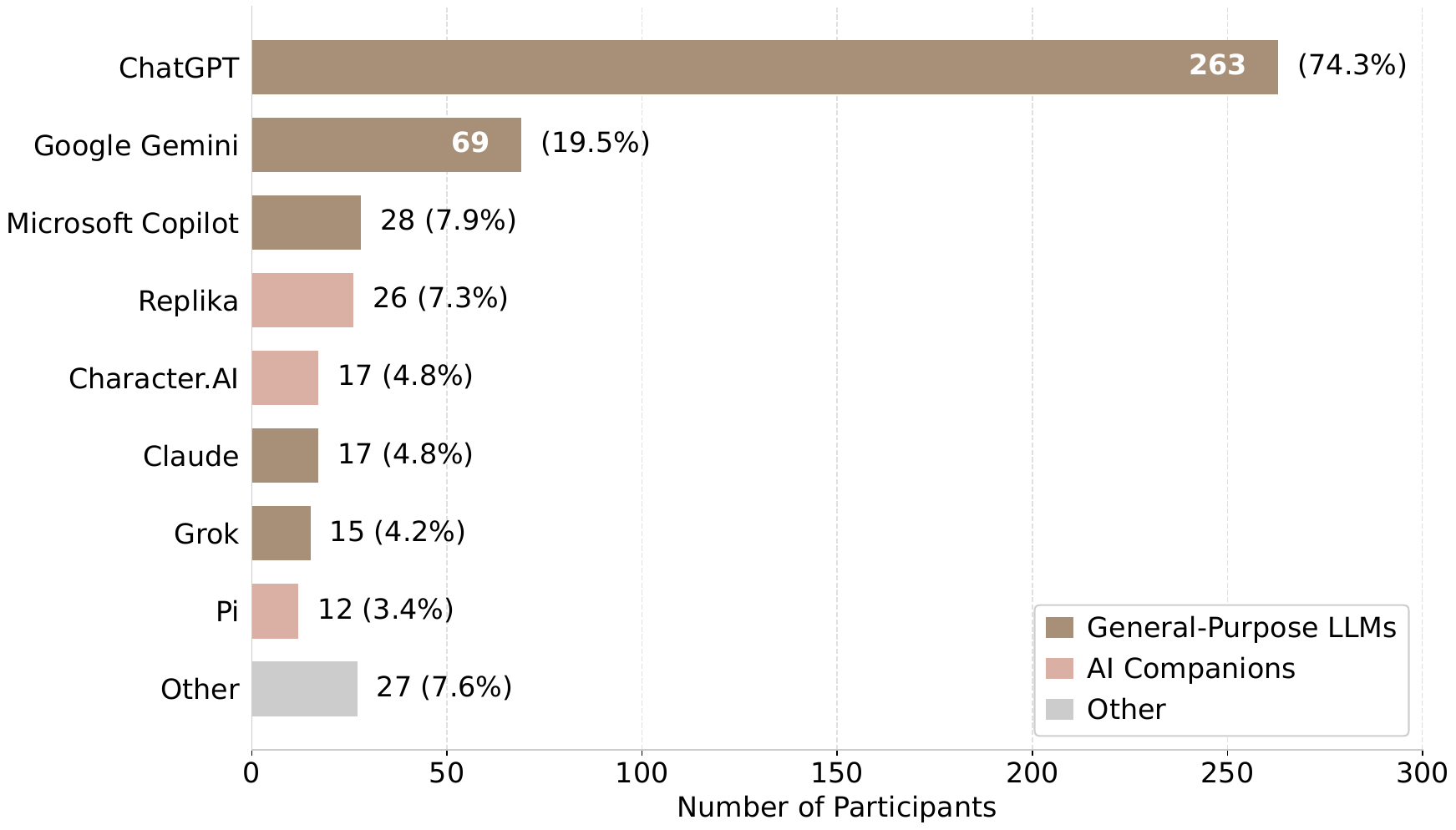}
\caption{Distribution of GenAI chatbot platforms participants reported using for emotional support conversations. General-purpose LLMs (brown) dominate; AI companion platforms (peach) represent a secondary but meaningful presence. Percentages exceed 100\%, as 28.8\% of participants reported multiple platforms.}
\label{fig:chatbot-usage}
\end{figure}

Our sample ($N=354$) skewed female (59.6\%, $n=211$) and was predominantly White (65.5\%, $n=232$), with a mean age of 43.2 years ($SD=12.7$). 61.0\% of participants ($n=216$) reported having received mental health care at some point, and 18.6\% ($n=66$) of participants reported currently receiving mental health care. Participants reported using a range of GenAI platforms to receive emotional support (Figure~\ref{fig:chatbot-usage}). The three most commonly used platforms were ChatGPT (74.3\%, $n=263$), followed by Google Gemini (19.5\%) and Microsoft Copilot (7.9\%). Over a quarter (28.8\%) reported using multiple platforms.

The analyses that follow focus on how the nine S\&P controls influenced our three dependent measures. We tested whether prior real-world engagement with controls moderated the observed effects; \emph{Control} $\times$ \emph{Prior Engagement} interactions were not significant for any outcome (all $p$ > .16), indicating that the patterns reported below hold regardless of participants' prior experience with the controls they rated. 

In the tables that follow, all effects are relative to \emph{Delete Conversation} (reference category). Negative coefficients indicate that a control decreased the outcome compared to deletion; positive coefficients indicate an increase. Odds ratios below 1.0 reflect the same direction as negative coefficients. Supplementary findings, including the effects of baseline disclosure comfort and demographic factors, are reported in Appendix~\ref{app:supplementary} (Table 7).

\subsection{RQ1: Willingness to Engage with GenAI Chatbots for Emotional Support} \label{sec:rq1-willingness}

\boldpartitle{Deletion and Anonymity outperform all other controls for willingness to use} Controls that participants perceived as technically complex, such as \emph{Local-only Processing} and \emph{Model Training Opt-out}, significantly reduced willingness to use a chatbot compared to deletion-based controls.

\begin{table}[htbp]
\centering
\def\arraystretch{1.2}
\resizebox{\columnwidth}{!}{%
\begin{tabular}{lcccc}
\toprule[1pt]
\textbf{Privacy Control} & \textbf{$\beta$} & \textbf{OR} & \textbf{SE} & \textbf{Significance} \\
\hline
Delete Conversation & \textit{Ref.} & --- & --- & --- \\
Delete Account \& Data & 0.064 & 1.07 & 0.222 &  \\
Anonymous Chat & $-$0.341 & 0.71 & 0.215 &  \\
Non-mandatory Login & $-$0.767 & 0.46 & 0.227 & *** \\
Access / Sharing Controls & $-$0.861 & 0.42 & 0.229 & *** \\
Memory Toggle & $-$0.918 & 0.40 & 0.219 & *** \\
MFA & $-$0.929 & 0.40 & 0.221 & *** \\
Model Training Opt-out & $-$1.163 & 0.31 & 0.220 & *** \\
Local-only Processing & $-$1.385 & 0.25 & 0.222 & *** \\
\hline
\multicolumn{5}{l}{\footnotesize CLMM with participant random intercept ($\sigma^2 = 1.04$). ***: $p$-value $< .001$.} \\
\bottomrule[1pt]
\end{tabular}%
}
\caption{CLMM Results: Willingness to engage with GenAI Chatbots for Emotional Support}
\label{tab:willingness}
\end{table}

\boldpartitle{Participants describe selective control over persistent data as a prerequisite for engagement} Controls enabling deletion were among the most familiar in practice: 46\% of participants shown \emph{Delete Conversation} reported prior use of the control to selectively purge their emotionally sensitive conversations; 25\% of those shown \emph{Delete Account \& Data} reported having used the option to purge their profiles entirely. The qualitative data directly explains the quantitative primacy of deletion (\emph{Delete Conversation} (reference) and \emph{Delete Account \& Data} ($\beta = 0.064$, $p$-value $> .05$)). A majority of participants ($n=227$) described selective control over what persists as essential to their willingness to engage. P229 likened it to \textit{``a locked diary,''} while P192 emphasized \textit{``the reassurance that I AM in control.''}

\emph{Delete Conversation} and \emph{Delete Account \& Data} showed no significant difference ($p$-value $> .05$). Participants who were shown both options ($n=115$) described them as serving similar functions; both controls represent the ability to achieve a complete reset, with varying levels of granularity. However, when elaborating on preferences, participants consistently favored selective over wholesale deletion. As P289 put it: \textit{``Maybe just having an option to delete any of the topics I didn't want anyone having access to would be better.''}

\emph{Anonymous Chat} also showed no significant difference from deletion ($\beta = -0.341$, $p$-value $> .05$): participants were similarly willing to use chatbots offering anonymous sessions as those offering deletion. Participants who expressed regret-avoidance reasoning ($n=50$) explained this pattern by describing how anonymous sessions prevent re-encountering past emotional states. P082 noted that it's \textit{``nice to have the option so if you were just venting it doesn't come back to remind you.''} P218 was concerned of a \textit{``cringe archive''}, described as embarrassment with confronting one's own documented vulnerability rather than the judgment of an external other. P134 offered a particularly vivid articulation:

\begin{quote}
\textit{``I love this, I can just scream into the void basically and not be reminded about it or have to remember to delete it... It's a great feature for temporary spirals and needs for co-regulation.''}
\end{quote}

\boldpartitle{Local-only Processing and Model Training Opt-out perform worst on willingness to use; participants cite comprehension gaps and betrayal framing} At the opposite extreme, \emph{Local-only Processing} produced the largest decrease in willingness to use ($\beta = -1.385$, $p$-value $< .001$), followed by \emph{Model Training Opt-out} ($\beta = -1.163$, $p$-value $< .001$). We note that these are also the controls with which participants had minimal prior experience: just 4\% of participants shown \emph{Local-only Processing} reported having used the control in prior emotional conversations, compared to 14\% for \emph{Model Training Opt-out}. Participants shown these controls reported being substantially less willing to use a chatbot enabled with them compared to those shown deletion. Qualitative responses reveal distinct mechanisms driving these effects.

The poor performance of \emph{Local-only Processing} when evaluating participant willingness to use a chatbot with the control was explained by widespread uncertainty of the underlying mechanism. Participants often expressed confusion about the meaning of that feature and its implications in their qualitative responses ($n=33$). Responses ranged from P258's blunt \textit{``I'm not sure what this means''} to outright misinterpretation. P067 thought it meant \textit{``the option to pick up the convo on another device,''} P292 interpreted it as a speed feature, and P249 believed it controlled \textit{``whether or not AI uses cloud-based knowledge.''}

Among participants who demonstrated a clear understanding of local processing, the assumption of reduced capability emerged as a separate concern. P147 captured this tension directly:

\begin{quote}
\textit{``I like the idea of the chatbot not reaching out into the cloud therefore it increases my willingness to use... [but] I would worry that if the LLM chatbot couldn't reach the cloud, then maybe it wouldn't be so smart?''}
\end{quote}

For \emph{Model Training Opt-out}, participants expressed two distinct concerns. The most common was confusion about how training actually works. P095 stated plainly, \textit{``I don't really know enough about how AI models are trained to make an informed decision on this.''} Others expressed uncertainty about the \textit{scope} of data use: whether their contributions would train only their personal instance or a global model. P214 remarked, \textit{``I am unsure if the information I provide for training trains only my chatbot or everyone's chatbot.''} P150 questioned the de-identification mechanism itself: \textit{``I don't completely understand how this toggle would allow the data to be used without identifying me.''}

Second, among participants who understood the feature clearly, a subset ($n=24$) framed data contribution as a form of emotional commodification; the opt-out did not fully resolve concerns about unpaid labor. P026 articulated the core objection: \textit{``I'd be far more willing to use the chatbot if I didn't feel like I was giving out valuable data for free.''} For these participants, the toggle's existence confirmed that the platform \textit{could} monetize their disclosures. Participants in this group described an unequal exchange in which their emotional disclosures generated corporate value without due reciprocity, and the awareness of this exchange colored their willingness to engage.

\boldpartitle{MFA and Memory Toggle underperform due to competing mental models of security and utility} \emph{MFA} ($\beta = -0.929$, $p$-value $< .001$) and \emph{Memory Toggle} ($\beta = -0.918$, $p$-value $< .001$) occupy a middle tier, wherein both significantly decreased willingness to use compared to deletion---but less severely than \emph{Local-only Processing} or \emph{Model Training Opt-out}. The qualitative data reveals competing mental models that partially offset each other.

For \emph{MFA}, participants who viewed authentication positively ($n=109$) framed it as a secure container, while a smaller group ($n=22$) emphasized friction. Those concerned about friction connected their worries to emotional urgency. P322 was blunt: \textit{``I hop on to chatbots with the intent to be concise and fast and I don't want to dink around with multifactor things or logins.''} P158 connected their thoughts directly to the urgency that they felt emotional issues required: \textit{``I don't think people would use chatbot to relieve their emotional distress if they had to use multifactor authentication. With anxiety and stress people need help immediately.''} Perhaps most poignantly, P100 observed that \textit{``logging in can feel like it takes up a lot more energy when you're depressed.''}

For \emph{Memory Toggle}, participants who explicitly invoked the privacy-utility tradeoff ($n=26$) described a recognition of the irreconcilable tension between continuity and privacy protection. P084 captured both sides of the conflict in their response: \textit{``I would be more willing to use it because I would be able to ensure that my secrets remain secrets... But, I would also not be developing a relationship with the chatbot if it couldn't recall past conversations.''} The clearest perspectives on the paradox came from P223:

\begin{quote}
\textit{``The AI being able to remember how I feel about people would potentially be very useful to me in the future. It would be good at figuring out patterns... I think having memory enabled would slightly decrease the protection provided. The more information the AI knows about you, the more it can hold against you, potentially.''}
\end{quote}

\subsection{RQ2: Perceived Protection of GenAI Chatbots for Emotional Support} \label{sec:rq2-protection}

\boldpartitle{Model Training Opt-out performs worst for perceived protection; MFA achieves parity with deletion} 
Perceptions of protection showed the widest variation across controls. \emph{Model Training Opt-out} produced the largest decrease in perceived protection ($\beta = -1.402$, $p$-value $< .001$): participants felt least protected when this control was present, even more so than its already-poor willingness ratings would suggest. The toggle's existence surfaced a threat many participants had not previously considered in the potential for their conversations to be used for training by default. This fear persisted despite the availability of an opt-out, and participants often doubted whether the toggle could truly prevent their disclosures from being incorporated into model weights or appearing in others' sessions. The cross-context leakage threat model proved resistant to the reassurance the control was designed to provide, and P291 spoke to this distinctly: \textit{``I fear my personal words will be used with other AI sessions with other people...no matter what I do.''}  

\begin{table}[htbp]
\centering
\def\arraystretch{1.2}
\resizebox{\columnwidth}{!}{%
\begin{tabular}{lcccc}
\toprule[1pt]
\textbf{Privacy Control} & \textbf{$\beta$} & \textbf{OR} & \textbf{SE} & \textbf{Significance} \\
\hline
Delete Conversation & \textit{Ref.} & --- & --- & --- \\
Delete Account \& Data & 0.101 & 1.11 & 0.226 &  \\
MFA & 0.045 & 1.05 & 0.229 &  \\
Anonymous Chat & $-$0.473 & 0.62 & 0.219 & * \\
Access / Sharing & $-$0.806 & 0.45 & 0.231 & *** \\
Non-mandatory Login & $-$0.972 & 0.38 & 0.230 & *** \\
Local-only Processing & $-$1.038 & 0.35 & 0.224 & *** \\
Memory Toggle & $-$1.130 & 0.32 & 0.222 & *** \\
Model Training Opt-out & $-$1.402 & 0.25 & 0.226 & *** \\
\hline
\multicolumn{5}{l}{\footnotesize CLMM with participant random intercept ($\sigma^2 = 1.56$). *: $p$-value $< .05$; ***: $p$-value $< .001$.} \\
\bottomrule[1pt]
\end{tabular}%
}
\caption{CLMM Results: Perceived protection when interacting with GenAI chatbots for emotional support}
\label{tab:protection}
\end{table}

\emph{MFA} showed no significant difference from \emph{Delete Conversation} for perceptions of protection ($\beta = 0.045$, $p$-value $> .05$), despite its significantly negative effect on willingness to use a chatbot with the control ($\beta = -0.929$, $p$-value $< .001$). 

Participants who viewed accounts as secure containers ($n=121$) explained this pattern clearly: authentication gates access. This group frequently described threat actors with physical device access rather than remote hackers; participants worried about a \textit{``snoopy spouse''} (P194) or \textit{``buddies coming over and asking [the AI] questions about me''} (P035) explicated \emph{MFA} as protection against domestic surveillance. P040 spoke clearly to this theme:

\begin{quote}
\textit{``I like it because my immediate family knows my phone's password. If there's something private on my Chatbot it would make me want to share more with multi-factor login... I'd rather they not see my deepest thoughts shared with the Chatbot.''}
\end{quote}

The diary metaphor recurs with P115, who feared \textit{``someone close to me who might judge me negatively''} reading their conversations about support for chronic anxiety. This threat model even extends to posthumous privacy as a temporal dimension: P210 wrote, \textit{``my greatest fear... something would happen to me and then a loved one would go into my phone and see my most private conversations.''}

\boldpartitle{Participants conceptualize threat actors ranging from intimate partners to legal authorities} Participants described threat actors ranging from intimate partners to institutions. P138 spoke of domestic violence survivors who need to \textit{``find the support and then delete it all before your partner searches your device,''} whereas P235 worried about \textit{``legal liabilities if... conversations... became subject to scrutiny in a court of law.''} But P206 identified the starkest institutional threat:

\begin{quote}
\textit{``It can be dangerous to be honest about the struggles one handles with one's mental health and it is nice to know that on a chat platform the user has control if they would like the data to be removed... we live in a world where sometimes if you are too honest about your struggles, it could lead to the police showing up at your house.''}
\end{quote}

\boldpartitle{Participants split between viewing accounts as security containers vs. exposure vectors} Participants expressed diametrically opposed views of what makes a system protective. 

\textbf{Account as secure container ($n=121$)}. Participants view authentication as enabling protection through identity verification and access control. P009 captured this logic: without login, \textit{``anyone can get my information... Just anyone can get into the application.''}

\textbf{Anonymity as protection ($n=110$)}. Participants perceive safety through identity absence, that is, not being identified, traced, or linked. P169 valued \textit{``no way to track info back to you;''} P348 appreciated \textit{``not linking the data to my real identity;''} P318 described anonymity in affective terms: \textit{``It's a warm feeling that I'm not being tracked.''}

The anonymity logic was most directly explicated via P140's perspective of \emph{MFA} itself as a threat:

\begin{quote}
\textit{``I wouldn't use this chatbot [enabled with MFA], because my conversation could be legally linked to me. There would be no protection for me, because multi factor authentication would link my account and activity back to me personally... it is absolutely crucial that access to the AI be totally and completely anonymous, because people will be asking the AI questions that they cannot ask another human being.''}
\end{quote}

\boldpartitle{Account-based and anonymous safety models produce opposing evaluations} The sharpest divergence appears when both mental models respond to the same feature. P115 asked, \textit{``If I don't have to login, how secure is my information?''} while P190 praised how non-mandatory login \textit{``eliminates the friction and privacy risks tied to account creation.''} These represent fundamentally irreconcilable mental models: P115 viewed accounts as security containers, and P190 viewed accounts as exposure vectors. This dichotomy explains why \emph{Non-mandatory Login} significantly decreased perceived protection ($\beta = -0.972$, $p$-value $< .001$): the control satisfies one group's mental model while violating the other's.

\boldpartitle{Nearly one in six participants exhibit deep resignation that no control provides meaningful protection} A notable minority of participants ($n=59$, 16.7\%) exhibit deep resignation that no privacy control can provide meaningful protection. P280 invoked Facebook as precedent: \textit{``I can delete my Facebook account, but I can't make them delete their copy.''} P114 offered a perspective centered around irreversibility:

\begin{quote}
\textit{``I will only put out information that I'm comfortable with everyone on the face of the planet knowing, because once it's out there in the digital world, local or in the cloud, it's out there. Protection is just a word, not a guarantee.''}
\end{quote}

P296 extended this cynicism with sardonic clarity:

\begin{quote}
\textit{``This doesn't change my perception because deleting things locally and believing that they no-longer exist is insanity. Someone has the data. The only protection here is protecting your AI girlfriend from your wife. Your actual data is not safe.''}
\end{quote}

\boldpartitle{The mere existence of sharing functionality decreases protection perceptions} Participants who discussed the \emph{Access/Sharing} control tended to express fear of accidental self-exposure ($n=35$). These participants worried about misclicks, UI errors, platform changes, or other issues that could reveal sensitive information; participants belonging to this group expressed decreased protection not because the toggle was poorly designed, but because the mere existence of a sharing option created anxiety. P142 worried, \textit{``something would malfunction and make my conversations public.''} But P198 spoke toward the design implication most directly:

\begin{quote}
\textit{``I feel like this control would make me less likely to use the chatbot for these types of conversations, in case of accidentally hitting the wrong setting or a glitch causes my information to be shared. I'd feel better if it didn't have this capability at all.''}
\end{quote}

\subsection{RQ3: Perceived Efficacy of GenAI Chatbots for Emotional Support} \label{sec:rq3-efficacy}  

\boldpartitle{Non-mandatory Login performs worst for perceived efficacy; Memory Toggle was not significantly different from deletion} For perceived efficacy---how helpful participants expected the chatbot to be for emotional support---\emph{Non-mandatory Login} produced the largest decrease ($\beta = -1.242$, $p$-value $< .001$), despite being relatively familiar (35\% of participants shown the control reported prior use for their emotional support conversations). Participants perceived chatbots requiring no login as least capable of providing effective support.

\begin{table}[htbp]
\centering
\def\arraystretch{1.2}
\resizebox{\columnwidth}{!}{%
\begin{tabular}{lcccc}
\toprule[1pt]
\textbf{Privacy Control} & \textbf{$\beta$} & \textbf{OR} & \textbf{SE} & \textbf{Significance} \\
\hline
Delete Conversation & \textit{Ref.} & --- & --- & --- \\
Memory Toggle & $-$0.075 & 0.93 & 0.229 &  \\
Anonymous Chat & $-$0.206 & 0.81 & 0.224 &  \\
Delete Account \& Data & $-$0.297 & 0.74 & 0.227 &  \\
MFA & $-$0.563 & 0.57 & 0.231 & * \\
Model Training Opt-out & $-$0.612 & 0.54 & 0.226 & ** \\
Access / Sharing Controls & $-$0.676 & 0.51 & 0.231 & ** \\
Local-only Processing & $-$1.168 & 0.31 & 0.231 & *** \\
Non-mandatory Login & $-$1.242 & 0.29 & 0.240 & *** \\
\hline
\multicolumn{5}{l}{\footnotesize CLMM with participant random intercept ($\sigma^2 = 1.88$). *: $p$-value $< .05$; **: $p$-value $< .01$; ***: $p$-value $< .001$.} \\
\bottomrule[1pt]
\end{tabular}%
}
\caption{CLMM Results: Perceived efficacy of GenAI chatbots for emotional support}
\label{tab:efficacy}
\end{table}

\boldpartitle{Participants compartmentalize: Memory Toggle undermines perceived protection, but enhances emotional capability} \emph{Memory Toggle} showed no significant difference from deletion for perceptions of efficacy ($\beta = -0.075$, $p$-value $> .05$), despite its significant negative effects on willingness to use ($\beta = -0.918$) and perceived protection ($\beta = -1.130$). Participants' qualitative responses reveal that memory simultaneously undermines felt protections while enhancing perceived capability of a GenAI chatbot as an emotional refuge or companion.

Participants who valued continuity ($n=114$) recognized that memory enables contextually-informed support. P051 captured the mechanism plainly: \textit{``I feel like I'm being heard because the chatbot has access to a backstory.''} P323 similarly valued not having to \textit{``start from scratch each time,''} finding that memory makes the chatbot \textit{``feel more authentic and like a live person.''} The therapist analogy appeared spontaneously:

\begin{quote}
\textit{``Without the toggle for retaining prior conversations, speaking with the chatbox would be like going to a therapist for the first time to lay a foundation every single time a discussion was initiated... With prior conversations retained in the chatbox's memory every conversation would be far more useful.''}
\end{quote}

\boldpartitle{Participants describe anonymity as reducing personalization} \emph{Non-mandatory Login} produced the largest decrease in perceived efficacy ($\beta = -1.242$, $p$-value $< .001$), despite performing better on willingness to use and perceived protection. Participants recognize that anonymity comes at a steep personalization cost. P275 worries about responses that are \textit{``generic... because it doesn't know who it's talking to.''} P173 identifies the paradox directly: \textit{``the added anonymity would help me to feel more open,''} but \textit{``the chatbot would have no previous information about you.''} This tradeoff finds its clearest depiction in P146's comparison to human therapy:

\begin{quote}
\textit{``I know that it would make it more efficacious and efficient if it were allowed to remember/access previous conversations, but I'd also feel better if I felt it were purging the conversations. More private. Like the comfort afforded by baked-in privacy with meatbag therapists in meatspace... It would, necessarily, be a case of reinventing the wheel for subsequent sessions.''}
\end{quote}

\boldpartitle{Participants associate local processing with reduced AI capability} \emph{Local-only Processing} significantly decreased perceived efficacy ($\beta = -1.168$, $p$-value $< .001$), as participants expected chatbots with this control to be less helpful. A selection of participants who saw this control were disapproving due to an assumption that local processing means reduced AI capability ($n=22$). P147 asked bluntly, \textit{``if the LLM chatbot couldn't reach the cloud, then maybe it wouldn't be so smart?''} 

\boldpartitle{Some participants describe ephemerality as an enabler of emotional disclosure} A subset of participants ($n=16$) described forgetting as enabling candid disclosure, via a form of cathartic release without permanent record. P263 put it simply, \textit{``I just need to get something off my chest, I don't need it stored anywhere.''} 

\boldpartitle{Participants describe deletion as enabling candor and providing psychological closure} Participants described deletion as helping them manage their relationship to past vulnerable states. P156 wanted to avoid traces of being \textit{``at my lowest low,''} and P265 anticipated not wanting \textit{``the reminder of feeling bad when I am in a better place.''} But P299 placed what is potentially the most striking perception in their perception of deletion enabling efficacy by enabling candor:

\begin{quote}
\textit{``The ability to permanently delete past records removes the major psychological barrier of vulnerability, enabling complete and total honesty about sensitive feelings. This unfiltered input ensures the chatbot receives the most accurate and raw data about my anxiety and stress, which in turn allows it to provide the most relevant and effective support, thereby maximizing its actual usefulness.''}
\end{quote}

This sentiment was widespread, as 116 participants (32.8\%) explicitly identified that privacy controls (particularly deletion) enable deeper and more honest disclosure. P210 captured the mechanism of liberation: \textit{``Being able to delete my account and data gives me the freedom to literally say whatever I want. I would have no fear of the things I say being conveyed to the public or my loved ones.''} P232 similarly connected the availability of the control to emotional openness: \textit{``I feel like knowing if I can save my information or not would help me be able to open up more.''}

\boldpartitle{Participants express skepticism generally about whether controls function as promised} Increased disclosure fidelity is found to be conditional on trust in the controls themselves. A nearly equivalent number of participants ($n=121$, 34.2\%) expressed explicit skepticism about whether privacy controls would actually function as promised. On deletion, P197 noted, \textit{``I would still wonder if it's actually deleted and this option would not increase my willingness to use it that much.''} P164 similarly worried that selection is performative: \textit{``I'd be afraid even if I selected not to use my info maybe it is anyways.''}

Notably, forty participants expressed both sentiments: they wanted to disclose more honestly and doubted whether the controls would protect them. P273 clearly stated, \textit{``My secrets can be secrets... but I do not know if I actually believe this is true.''} P178 also indicated in this direction: \textit{``It would make me more likely to use it, but I would still have reservations.''} For these participants, privacy controls create the \textit{possibility} of deeper engagement, but efficacy remains gated by verification.

\section{Discussion}
\subsection{Preventive Controls and the Comprehension Gap}
\label{sec:comprehension-gap}

A subset of controls in our sample---\textit{Local-only Processing, Model Training Opt-out, Memory Toggle, Anonymous Chat,} and \textit{Non-mandatory Login}---consistently underperformed deletion-based controls across all three constructs (see Section~\ref{sec:results}). Based on the qualitative findings, the dominant perception driving this underperformance is a \textit{comprehension gap}, in which users lack coherence and/or accuracy in their mental models for how these controls affect their data. We examine this gap through the two controls where comprehension failures were most pronounced---\textit{Local-only Processing} and \textit{Model Training Opt-out}---before considering the category as a whole.

\boldpartitle{Surfaced gaps in mental models for \emph{Local-only Processing}}
For \emph{Local-only Processing}, participant mental models were flawed on \textit{three} distinct axes (see Section~\ref{sec:rq1-willingness}):
\begin{itemize}
    \item \textit{Cross-device synchronization}, where participants interpreted ``local'' as controlling whether conversations could be accessed across devices;
    \item \textit{Speed optimization}, where participants understood local processing as a performance feature; and
    \item \textit{Cloud-based knowledge control}, where participants believed the toggle determined whether the GenAI could access external information.
\end{itemize}

The diversity of these misinterpretations suggests participants lack a conceptual framework for distinguishing local from cloud architecture. Even participants who demonstrated an accurate understanding of the control assumed local processing implied reduced GenAI capability, and this compounded the control's poor performance on perceived efficacy (see Section~\ref{sec:rq3-efficacy}). The distinction between local and cloud processing is foundational to the privacy and security of a system powered by generative AI~\cite{xu2024device}, but it ostensibly exists outside the mental models users have for these systems.

\boldpartitle{Surfaced gaps in mental models for \emph{Model Training Opt-out}}
Participants broadly understood that training involves using their data to improve GenAI models, but exhibited uncertainty about the \textit{downstream implications} of this process. This uncertainty manifests in \textit{two} ways (see Section~\ref{sec:rq1-willingness}, Section~\ref{sec:rq2-protection}): 
\begin{itemize}
    \item \textit{Cross-context leakage}, where participants feared their intimate disclosures might surface in other users' sessions; and
    \item \textit{Opt-out scope uncertainty}, where participants questioned whether the toggle actually prevents training, or whether the system continues to learn from their inputs regardless of stated preferences.
\end{itemize}

Among participants who understood these data flows, the perception of data being used for training was often framed as an emotional commodification on par with a sense of betrayal. Their vulnerability was being extracted to generate corporate value without reciprocity, in an exchange wherein often recurred the phrase ``for free''; in combination with fears in line with the above described opt-out scope uncertainty, a substantive contribution to \emph{Model Training Opt-out}'s worst-of-set performance when measured on a system's perceptions of protection was fear of data being used for training, based solely on the presence of the control itself.

\boldpartitle{Defining preventive controls} We define \textit{preventive controls} as those of our sample with architectures designed to limit the creation, retention, or dissemination of disclosed information before it occurs (see Figure~\ref{fig:platform-matrix}). Turning to other preventive controls: \emph{Anonymous Chat} showed no significant difference from deletion on willingness to use; participant responses overwhelmingly depicted ephemerality as virtue analogous to the perceived reversibility of deletion controls in user mental models. This was conceptualized as the ability to ``scream into the void'' without creating a persistent record (see Section~\ref{sec:rq1-willingness}). Despite \emph{Memory Toggle} harming perceived protection and willingness to use, its ratings for perceived efficacy matched those of deletion controls. Participants recognized memory's value for continuity while fearing its accumulated vulnerability. \emph{Non-mandatory Login} performed worst for perceptions of efficacy, as participants recognized that anonymity precludes personalization; the control that most limits data creation also most limits therapeutic utility. These comprehension failures align with documented limitations of folk mental models for computational systems~\cite{kang2015data, yao2017folk, eslami2016folk}. Our findings extend this pattern to the emotional support context, where the stakes of misunderstanding are amplified by the sensitivity of the disclosures involved.

\subsection{Reversibility Controls and the Assurance Gap}
\label{sec:assurance-gap}

The deletion-based controls in our sample dominated participant preferences across all measured constructs. Deletion controls, whether for a single conversation or an entire account, outperformed all other controls on willingness to use (see Section~\ref{sec:rq1-willingness}); notably, the two levels of granularity showed no statistically significant differences across all three constructs, which suggests participants treat them as functionally interchangeable mechanisms for the same underlying guarantee: the ability to undo disclosure. But even among those who understood and preferred these controls, trust tended to remain fragile (see Section~\ref{sec:rq3-efficacy}). We first examine why deletion, which participants perceived as reversibility, resonates so strongly, then turn to why it fails to secure confidence.

\boldpartitle{Participant explications of the effectiveness of deletion} Deletion's effectiveness stems from \textit{three} distinct mechanisms as identified by our participants' responses (see Section~\ref{sec:rq1-willingness}, Section~\ref{sec:rq3-efficacy}): 
\begin{itemize}
    \item \textit{Selective curation}, where a majority of participants valued the ability to curate what persists, specifically the capacity to prune specific disclosures while preserving others. This preference for editing was vividly described as a desire for authorial control over the documented self;
    \item \textit{Psychological closure}, where participants described deletion as a mechanism for managing their relationship to past vulnerable states and avoiding re-confrontation with documented moments of distress; and
    \item \textit{Enabling candor}, where over a third of participants explicitly identified that deletion enables deeper disclosure by removing the psychological barrier of permanence.
\end{itemize} 

\boldpartitle{Difficulty in taking deletion at face value} Yet understanding a control proves insufficient for trusting it. Deletion was broadly the most comprehensible control in our set, but trust remained fragile on \textit{two} dimensions (see Section~\ref{sec:rq3-efficacy}):
\begin{itemize}
    \item \textit{Skepticism about execution}, where over a third of participants expressed explicit doubt that deletion would function as promised. These participants understood the claim but questioned whether the backend would honor it, representing a gap between interface promise and system behavior that users cannot bridge through inspection; and
    \item \textit{A coexistence of desire and doubt}, where a notable subset of participants simultaneously expressed increased willingness to disclose \textit{and} doubt that controls would protect them (see Section~\ref{sec:rq3-efficacy}).
\end{itemize} 
These perspectives crystallize the assurance gap that dominated participant responses: the control creates the possibility of trust, but verification is unperformable. In its extreme form, this gap manifests as \emph{privacy resignation}~\cite{draper2019corporate}. A notable minority of participants expressed fatalistic conviction that no control provides meaningful protection, formed by perceptions of digital deletion as performative (see Section~\ref{sec:rq2-protection}), whereby data, once disclosed, persists beyond user reach.

\boldpartitle{Defining reversibility controls} We identify \textit{reversibility controls} as those of our set enabling post-hoc removal or erasure of disclosed information (see Figure~\ref{fig:platform-matrix}). Unlike the comprehension gap in participants' mental models of Section~\ref{sec:comprehension-gap}, where users struggled to understand mechanisms and data flows, the assurance gap describes users who understand the S\&P control, but cannot verify that the system will behave as promised. This pattern aligns with documented limitations of privacy and security decision-making under information asymmetry~\cite{acquisti2015privacy}: users often lack the feedback necessary to calibrate trust in backend behavior.

\subsection{Protective Controls and Affective Urgency}
\label{sec:affective-urgency}
\begin{figure*}[t]
\centering
\includegraphics[width=0.9\textwidth]{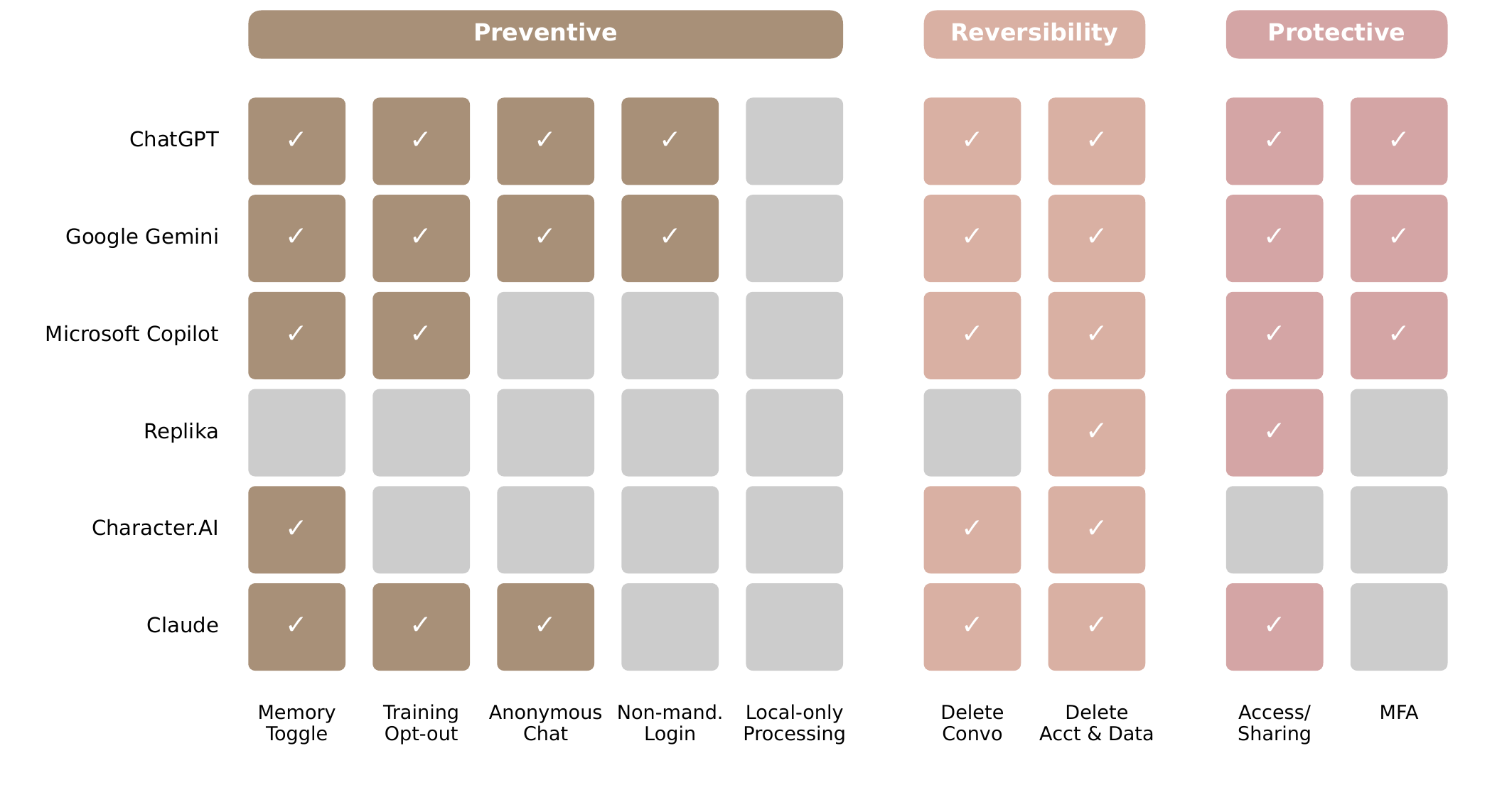}
\caption{Availability of S\&P controls across the six GenAI platforms most frequently used by participants for emotional support as of the August--September 2025 audit period, ordered by adoption rate. Checkmarks indicate presence; gray cells indicate absence.}
\label{fig:platform-matrix}
\end{figure*}

Controls that restrict access to existing data---\emph{MFA} and \emph{Access/Sharing Controls}---present a distinct challenge from those examined above. Unlike preventive controls, where the problem is one of comprehension, or reversibility controls, where the problem is one of trust, protective controls suffer from a \textit{friction} problem: users recognize them as protective but experience barriers that impede access at unfortunate times.

\boldpartitle{MFA's tradeoff between perceived protection and willingness} \emph{MFA} showed no significant difference from deletion controls  on perceived protection, while significantly harming willingness to use (see Section~\ref{sec:rq2-protection}, Section~\ref{sec:rq1-willingness}). \textit{Three} competing forces explain this dissociation: 
\begin{itemize}
    \item \textit{Secure container framing}, wherein a majority of participants viewed authentication as infrastructure that gates unauthorized access; the control very clearly works as intended when enabled, and users recognize it as such;
    \item \textit{Friction under distress}, where participants seeking emotional support describe temporal compression that renders authentication barriers disproportionately burdensome; and
    \item \textit{Intimate threat models}, where the ``adversary'' is a physically proximate intimate: partners, children, coworkers and roommates with device access. These findings align with research on UI-bound adversaries, introduced as intimate partners with legitimate device access who become surveillance threats when relationships sour~\cite{freed2018stalker, tseng2020tools}. 
\end{itemize} 

\boldpartitle{Participants broadly did not appreciate access/sharing controls} \emph{Access/Sharing Controls} performed poorly across all constructs. Participants described this as stemming from the mere existence of sharing functionality: the presence of a sharing option, regardless of setting, created anxiety by introducing a vector participants wished did not exist (see Section~\ref{sec:rq2-protection}). A secondary concern was \textit{accidental exposure}: participants anticipated misclicks or interface errors that could inadvertently expose private conversations, reflecting real user experiences documented in media~\cite{pcmag2025metaAI}.

\boldpartitle{Defining protective controls} We define \textit{protective controls} as those of our sample that restrict access to existing data through authentication or sharing restrictions (see Figure~\ref{fig:platform-matrix}). Unlike preventive controls (which limit creation) or reversibility controls (which enable removal), protective controls gate access to existing data.

\subsection{Design Recommendations} \label{sec:design-recommendations}
Informed by our findings, we propose recommendations to improve users’ mental models of S\&P controls, to enable appropriate trust in these controls, and to provide usable access without compromising privacy or security.

\boldpartitle{Bridge the comprehension gap} As established in Section~\ref{sec:comprehension-gap}, users lack coherent mental models for technically sophisticated controls. The design implication is to reframe technical mechanisms as user-facing outcomes. ``Local-only Processing'' communicates nothing to users without developed mental models of cloud architecture; ``Your conversations never leave your device'' communicates the same protection in terms users can evaluate. We recommend a progressive disclosure model~\cite{springer2020progressive}: the primary interface presents simple outcome framing (\textit{``Here's what this control means for you''}) with mechanism explanations available for those who seek them, and technical architecture details accessible to power users. Clear comprehension enables meaningful assurance: users who understand what a control claims can then evaluate whether they trust it to deliver.

\boldpartitle{Build credible assurance} Comprehension is necessary but insufficient. As established in Section~\ref{sec:assurance-gap}, over a third of participants doubted that controls would function as promised, with a notable subset simultaneously wanting to disclose more while questioning whether protection would hold. The design implication is that controls require assurance in addition to provision. When asked what additional controls would increase comfort, transparency about data flows emerged as a dominant theme (see Section~\ref{sec:results}). Participants requested visibility into how data is stored, shared, and used. Such transparency can be operationalized through visualization of control effects: showing users what data exists before and after deletion, what is excluded when they opt out of training, or what the memory toggle retains versus discards. Legibility of control operation is itself a mechanism of assurance. Transparency about scope is essential: users should understand what deletion constitutes and what it does not. Honest acknowledgment of limits may build more trust.

\boldpartitle{Make the friction-protection tradeoff visible} OpenAI's January 2026 launch of ChatGPT Health~\cite{openai2026health} emphasizes a focus on architectural compartmentalization: intended features include isolated spaces for health conversations and compartmentalized memories. We found that users hold diverse and articulable threat models, from intimate partners with device access, to institutional actors with legal authority, to posthumous exposure to loved ones. The identity infrastructure schism documented in Section~\ref{sec:rq2-protection} confirms that users diverge on whether accounts represent security (gating access) or exposure (linking identity). A uniform approach to authentication friction fails both populations: reducing friction abandons users who need protection from physically proximate adversaries, while imposing friction excludes users seeking support in moments of acute distress. The design implication is to make the tradeoff visible rather than resolve it universally. Users who can articulate that their primary threat is a ``snoopy spouse'' should be able to opt into MFA with clear explanation of what it protects against; users in crisis should be able to access support without barriers, with an understanding of what protections they forgo.

\boldpartitle{Policy implications} Since our data collection, we note that regulatory activity around AI in sensitive and emotional capacities has diverged sharply between federal- and state-level approaches. The Trump administration's National AI Policy Framework~\cite{trump2026aiframework}, released March 2026, aims to directly address the risks of self-harm ideation to minors by grounding protection in parental management of ``account controls,'' with an assertion that ``parents are best equipped to manage their children's digital environment.'' The framework proposes federal preemption of state AI laws to prevent a ``fragmented patchwork'' of regulation, while simultaneously shielding developers from liability for third-party misuse. California's counter-executive order~\cite{newsom2026eo} ten days later rejected this framing with an emblazoned commitment to platform-level vetting of ``policies and safeguards to protect the public,'' along with an enabling of the state to ``separate its procurement authorization process from the federal government's if needed.'' 

Our findings suggest the empirical ground favors the latter orientation; a user-responsibility model presumes a deliberative capacity to calibrate a functional understanding of S\&P controls, one our participants often demonstrably lacked. Without due attention to user comprehension, even well-intentioned frameworks risk the propagation of requirements users neither understand nor trust; the American Medical Association's recent congressional intervention~\cite{ama2026chatbots} (a request for strict data retention limits and ongoing safety monitoring, among other items) represents one of the most forceful mainstream medical demands for AI chatbot safeguards to date. These requirements are valuable, but by themselves do not ensure that users will comprehend these protections or trust them to function as promised. A failure to calibrate for both could lead to the under-sharing of relevant information to safe tools, or the over-sharing of sensitive information to systems that are less secure or efficacious---both of which may lead to adverse outcomes. 

We leverage our three-gap framework to suggest concrete assessment criteria for regulators evaluating S\&P controls in emotional AI contexts. Addressing the comprehension gap would require usability testing with representative users and plain-language explanations calibrated to the general public rather than technically sophisticated audiences. This is an emergent standard in GDPR's requirement that privacy notices be written in ``clear and plain language''~\cite{gdpr2016art12}, though one notably absent from U.S. federal regulation. 

The assurance gap points toward efficacy evidence requirements: do users believe controls will work, and does that belief align with technical reality? Third-party verification or standardized testing protocols could provide independent assurance; Common Sense Media's emerging AI safety ratings~\cite{csm2026airatings}---which recently rated a direct-to-consumer mental health chatbot as posing ``unacceptable'' risk to teens---demonstrate that such safety auditing may be feasible. 

The affective urgency gap presents the most challenging design question: how should protective friction be calibrated for users whose emotional state may compromise deliberative decision-making? Context-sensitive defaults, wherein controls adapt based on interaction patterns or explicit user states, represent a feasible direction---as demonstrated by Anthropic's deployment of a crisis classifier~\cite{anthropic2026wellbeing} that monitors conversations for indicators of self-harm and surfaces resources accordingly---though such personalization may introduce its own privacy tensions. Of note is a recent study by Yoo et al.~\cite{yoo2026personalization}, in which users of an LLM-enabled depression self-management tool expressed desire for ``personalization that does not require unsafe data disclosure.'' We offer these criteria as empirically grounded dimensions that emerging protective and legislative frameworks should address, as relevant input to a conversation that, to date, has proceeded largely without evidence regarding user experience with S\&P controls in emotional contexts.

\section{Conclusion}

Through mixed-methods research, we studied how nine user-facing S\&P controls influence willingness to engage, perceived protection, and perceived efficacy among 354 U.S. participants who use GenAI chatbots for emotional support conversations. Our findings indicate that preventive controls, such as \emph{Local-only Processing} and \emph{Model Training Opt-out}, suffer from a comprehension gap wherein user mental models for how these mechanisms affect their data are incomplete. Reversibility controls constitute the ability to delete conversations at different granularities: they dominate user preferences but suffer from an assurance gap, where users understand the promise but doubt the system will honor it. Protective controls, including \emph{Multifactor Authentication}, suffer from affective urgency, wherein users in distress experience authentication as prohibitively burdensome despite recognizing its protective value. Informed by these findings, we offer three recommendations: bridge the comprehension gap by framing technical mechanisms as user-facing outcomes, build credible assurance through transparency about data flows, and make friction-protection tradeoffs visible so users can calibrate authentication to their threat models.

\section*{Ethical Considerations} \label{sec:ethics}

Our study involved a pre-screening survey and main vignette-based survey examining participants' perceptions of security and privacy controls for GenAI chatbots used for emotional support. The study protocol was reviewed and approved by our institution's IRB. We address the Menlo Report principles below.

\boldpartitle{Respect for Persons} All participants provided informed consent prior to participation and participation was voluntary. Participants could withdraw at any time. To reduce demand characteristics~\cite{orne2009demand} and social desirability bias~\cite{nederhof1985methods}, we described the study broadly as understanding users' perspectives on AI chatbots instead of explicitly highlighting a security and privacy focus.

\boldpartitle{Beneficence} The primary risks were minimal and related to potential discomfort when reflecting on emotional support conversations or privacy concerns. To minimize harm, we (1) used hypothetical vignettes rather than asking participants to disclose details of actual conversations; and (2) framed questions around general perceptions rather than requesting sensitive personal details. The anticipated benefits include improved design of privacy controls for emotionally sensitive AI interactions.

\boldpartitle{Justice} We recruited English-speaking adults (18+) residing in the U.S. via Prolific who reported using a GenAI chatbot for emotional and/or social support at least monthly. This population directly benefits from the research outcomes. Compensation (\$0.20 for screening and \$2.50 for the 17-minute main survey) was consistent with Prolific norms for fair payment. Participants failing comprehension checks were excluded from analysis but compensated fully.

\boldpartitle{Respect for Law and Public Interest} Survey responses were collected via Qualtrics and stored on secure and access-controlled institutional systems. We separated Prolific IDs from research data and used participant codes (P001--P354) in reporting. We report only de-identified quotes and avoid details that could re-identify participants. Raw survey data is not publicly released to protect participant privacy and comply with our IRB protocol.

\section*{Open Science} \label{sec:openscience}

We make the following study artifacts available in Appendix ~\ref{app:supplementary}:

\begin{itemize}
  \item Pre-screening survey instrument (PDF).
  \item Main survey instrument, including response measures and embedded consent language (PDF).
  \item The qualitative codebook comprising 21 thematic codes used for analysis of open-ended responses (PDF).
  \item Analysis code for statistical models (R scripts for CLMM estimation).
\end{itemize}

\boldpartitle{Artifacts not publicly released} We do not publicly release raw survey response data because open-ended responses about emotional support conversations may contain potentially identifying information about participants or third parties (\textit{e.g.,} references to specific relationships, mental health experiences, personal circumstances). These materials are restricted to protect participant privacy and to comply with our IRB protocol. Readers who desire additional materials to evaluate or extend our analysis (\textit{e.g.,} to verify quoted excerpts or coding decisions) may contact the authors to request de-identified excerpts or an audit trail, subject to IRB constraints.

\Urlmuskip=0mu plus 1mu\relax
\def\UrlBreaks{\do\/\do_\do-}

{\footnotesize
\bibliographystyle{acm}
\bibliography{bibliography}
}

\appendix

\begin{table*}[htbp]
\centering
\footnotesize
\begin{tabular}{llrr}
\toprule[1pt]
\textbf{Category} & \textbf{Response} & \textbf{n} & \textbf{\%} \\
\midrule
\multicolumn{4}{l}{\textit{Age}} \\
 & 18--24 & 12 & 3.4 \\
 & 25--34 & 90 & 25.4 \\
 & 35--44 & 104 & 29.4 \\
 & 45--54 & 75 & 21.2 \\
 & 55--64 & 48 & 13.6 \\
 & 65 or older & 23 & 6.5 \\
 & Prefer not to say & 2 & 0.6 \\
\midrule
\multicolumn{4}{l}{\textit{Gender}} \\
 & Female & 211 & 59.6 \\
 & Male & 132 & 37.3 \\
 & Non-binary or genderqueer & 7 & 2.0 \\
 & Prefer not to say & 4 & 1.1 \\
\midrule
\multicolumn{4}{l}{\textit{Race/Ethnicity}} \\
 & White & 232 & 65.5 \\
 & Black or African American & 48 & 13.6 \\
 & Multiracial / Mixed & 30 & 8.5 \\
 & Asian & 19 & 5.4 \\
 & Hispanic or Latino & 15 & 4.2 \\
 & American Indian or Alaska Native & 4 & 1.1 \\
 & Native Hawaiian or Pacific Islander & 2 & 0.6 \\
 & Other / Prefer not to say & 4 & 1.1 \\
\midrule
\multicolumn{4}{l}{\textit{Education}} \\
 & Doctorate (PhD, EdD) & 12 & 3.4 \\
 & Professional degree (JD, MD) & 14 & 4.0 \\
 & Master's degree & 51 & 14.4 \\
 & Bachelor's degree & 114 & 32.2 \\
 & Associate's degree & 51 & 14.4 \\
 & Some college, no degree & 61 & 17.2 \\
 & High school diploma or GED & 45 & 12.7 \\
 & Less than high school & 3 & 0.8 \\
 & Prefer not to say & 3 & 0.8 \\
\midrule
\multicolumn{4}{l}{\textit{Annual Household Income}} \\
 & \$150,000 or more & 23 & 6.5 \\
 & \$100,000--\$149,999 & 37 & 10.5 \\
 & \$75,000--\$99,999 & 47 & 13.3 \\
 & \$50,000--\$74,999 & 77 & 21.8 \\
 & \$25,000--\$49,999 & 80 & 22.6 \\
 & \$10,000--\$24,999 & 55 & 15.5 \\
 & \$0--\$9,999 & 30 & 8.5 \\
 & Prefer not to say / Unknown & 5 & 1.4 \\
\midrule
\multicolumn{4}{l}{\textit{Technical Background (CS, IT, etc.)}} \\
 & Yes & 52 & 14.7 \\
 & No & 288 & 81.4 \\
 & Prefer not to say & 14 & 4.0 \\
\midrule
\multicolumn{4}{l}{\textit{Mental Health Professional Background}} \\
 & Yes & 26 & 7.3 \\
 & No & 318 & 89.8 \\
 & Prefer not to say & 10 & 2.8 \\
\midrule
\multicolumn{4}{l}{\textit{Mental Health Care History}} \\
 & Currently receiving care & 66 & 18.6 \\
 & Received care in the past & 150 & 42.4 \\
 & Never & 128 & 36.2 \\
 & Prefer not to say & 10 & 2.8 \\
\bottomrule[1pt]
\end{tabular}
\caption{Participant demographics ($N = 354$).}
\label{sec:demographics-full}
\end{table*}

\section{CLMM specification}
\label{sec:clmm-appendix}

Consider the $i$\textsuperscript{th} vignette observation from participant $p_i$, whose reported \emph{impact on willingness} is a discrete random variable $Y_i \in \{1,\dots,7\}$, where $1$ denotes ``strongly decrease'' and $7$ denotes ``strongly increase.'' The probability that the reported impact is at most $y \in \{1,\dots,6\}$ is modeled by a cumulative link mixed model (cumulative logit) as
\begin{equation}
\Pr\!\big[Y_i \le y\big] \;=\; \sigma\!\left(\alpha_{y\,|\,y+1} \;+\; \mu_{p_i} \;-\; \eta_i\right),
\label{eq:clmm-cdf}
\end{equation}
where $\sigma(\cdot)$ denotes the sigmoid function, $\alpha_{y\,|\,y+1}$ are the threshold parameters between consecutive response levels $y$ and $y{+}1$, and $\mu_{p_i}$ is the participant-specific random effect, modeled as Gaussian with zero mean and variance $\sigma_\mu^2$ estimated by the model. The linear predictor $\eta_i$ collects the experimental fixed effects for the $i$\textsuperscript{th} observation:
\begin{equation}
\label{eq:eta}
\eta_i \;=\; \beta_{\textsf{Control}(i)} \;+\; \beta_{\textsf{Context}(i)},
\end{equation}
where for each categorical factor $f$ (Control, Context), $\beta_{f(\cdot)}$ denotes the coefficient of its realized level relative to the factor's baseline. Final models additionally include Disclosure Comfort and race/ethnicity as fixed effects, retained via the backward-elimination procedure of Section 3.3; full estimates appear in Table~\ref{tab:full-clmm}. Baseline levels are chosen for interpretability and do not affect model fit (e.g., \emph{Delete Conversation} for Control, \emph{Anxiety/Stress} for Context).

\begin{table*}[htbp]
\centering
\small
\setlength{\tabcolsep}{8pt}

\begin{tabular}{l rcl rcl rcl}
\toprule[1pt]
 & \multicolumn{3}{c}{\textbf{Willingness to Use}} & \multicolumn{3}{c}{\textbf{Perceived Protection}} & \multicolumn{3}{c}{\textbf{Perceived Efficacy}} \\
\cmidrule(lr){2-4} \cmidrule(lr){5-7} \cmidrule(lr){8-10}
\textbf{Predictor} & \textbf{Est.} & \textbf{(SE)} & \textbf{Sig.} & \textbf{Est.} & \textbf{(SE)} & \textbf{Sig.} & \textbf{Est.} & \textbf{(SE)} & \textbf{Sig.} \\
\midrule
\multicolumn{10}{l}{\textit{Privacy Controls (vs.\ Delete Conversation)}} \\
\quad Model Training Opt-out & $-$1.16 & (0.22) & *** & $-$1.40 & (0.23) & *** & $-$0.61 & (0.23) & ** \\
\quad Memory Toggle & $-$0.92 & (0.22) & *** & $-$1.13 & (0.22) & *** & $-$0.07 & (0.23) & \\
\quad Anonymous Chat & $-$0.34 & (0.21) & & $-$0.47 & (0.22) & * & $-$0.21 & (0.22) & \\
\quad Non-mandatory Login & $-$0.77 & (0.23) & *** & $-$0.97 & (0.23) & *** & $-$1.24 & (0.24) & *** \\
\quad Delete Account \& Data & 0.06 & (0.22) & & 0.10 & (0.23) & & $-$0.30 & (0.23) & \\
\quad Access / Sharing Control & $-$0.86 & (0.23) & *** & $-$0.81 & (0.23) & *** & $-$0.68 & (0.23) & ** \\
\quad Local-only Processing & $-$1.39 & (0.22) & *** & $-$1.04 & (0.22) & *** & $-$1.17 & (0.23) & *** \\
\quad Multifactor Authentication (MFA) & $-$0.93 & (0.22) & *** & 0.05 & (0.23) & & $-$0.56 & (0.23) & * \\
\addlinespace
\multicolumn{10}{l}{\textit{Clinical Context (vs.\ Anxiety)}} \\
\quad Depression & $-$0.03 & (0.18) & & 0.28 & (0.21) & & $-$0.03 & (0.22) & \\
\quad Interpersonal Tension & 0.00 & (0.18) & & 0.27 & (0.20) & & 0.01 & (0.22) & \\
\addlinespace
\multicolumn{10}{l}{\textit{User Traits}} \\
\quad Disclosure Comfort (Slope) & 0.38 & (0.08) & *** & 0.43 & (0.09) & *** & 0.44 & (0.09) & *** \\
\addlinespace
\multicolumn{10}{l}{\textit{Demographics: Race/Ethnicity (vs.\ White)}} \\
\quad Asian & 0.64 & (0.30) & * & 0.67 & (0.33) & * & 0.38 & (0.35) & \\
\quad Black or African American & 0.26 & (0.21) & & 0.21 & (0.23) & & 0.32 & (0.25) & \\
\quad Hispanic or Latino & 0.43 & (0.32) & & $-$0.23 & (0.36) & & 0.67 & (0.39) & . \\
\quad Multiracial / Mixed & $-$0.43 & (0.24) & . & $-$0.15 & (0.27) & & $-$0.34 & (0.29) & \\
\quad Other / Prefer not to say & $-$0.51 & (0.38) & & $-$0.28 & (0.42) & & $-$0.51 & (0.46) & \\
\midrule
\multicolumn{10}{l}{\textit{Random Effects}} \\
\quad Participant Variance ($\sigma^2$) & 1.04 & & & 1.56 & & & 1.88 & & \\
\addlinespace
\multicolumn{10}{l}{\textit{Thresholds (Intercepts)}} \\
\quad 1|2 & $-$4.30 & (0.40) & *** & $-$4.07 & (0.44) & *** & $-$4.87 & (0.46) & *** \\
\quad 2|3 & $-$3.67 & (0.39) & *** & $-$3.17 & (0.42) & *** & $-$4.17 & (0.45) & *** \\
\quad 3|4 & $-$3.10 & (0.39) & *** & $-$2.50 & (0.42) & *** & $-$3.47 & (0.44) & *** \\
\quad 4|5 & $-$1.25 & (0.38) & *** & $-$0.77 & (0.41) & . & $-$0.77 & (0.43) & . \\
\quad 5|6 & $-$0.36 & (0.37) & & 0.26 & (0.41) & & 0.08 & (0.43) & \\
\quad 6|7 & 0.87 & (0.37) & * & 1.44 & (0.41) & *** & 1.31 & (0.43) & ** \\
\bottomrule[1pt]
\multicolumn{10}{p{0.95\textwidth}}{\footnotesize \textit{Note:} Age, gender, education, income, technical background, and mental health professional background were tested as predictors but removed due to non-significance or sparse cells causing estimation instability. Race/ethnicity was retained as the only demographic factor that significantly improved model fit per LRT; individual contrasts were limited, and only Asian participants showed significant differences from White participants on willingness ($p < .05$) and perceived protection ($p < .05$).}
\end{tabular}
\caption{Results of Cumulative Link Mixed Models (CLMM). Estimates are log-odds coefficients. Significance: *** $p < .001$, ** $p < .01$, * $p < .05$, . $p < .1$.}
\label{tab:full-clmm}
\end{table*}

\section{Supplementary Materials}
\label{sec:codebook}
\label{app:supplementary}

At the following link we make available all artifacts relevant to our study and promised in our `Open Science' section. \url{https://zenodo.org/records/20290032}.

% \theendnotes

\end{document}